\documentclass[floatfix,twocolumn,showpacs,preprintnumbers,amsmath,amssymb,prl,superscriptaddress]{revtex4}
\newcommand{\etal}{{\it et al.}}
\usepackage{graphicx}
\usepackage{dcolumn}
\usepackage{bm}

\begin{document}


\title{Strong Correlation Effects in the Fullerene C$_{20}$ }

\author{Fei Lin}
\affiliation{Department of Physics, University of Illinois at Urbana-Champaign, Urbana, Illinois 61801, USA}
\author{Erik~S.~S\o rensen}%
\author{Catherine Kallin}
\author{A. John Berlinsky}
\affiliation{Department of Physics and Astronomy, McMaster University, Hamilton,
ON, L8S 4M1 Canada }

\date{\today}

\begin{abstract}
The smallest fullerene, dodecahedral C$_{20}$, is studied using a one band
Hubbard model parameterized by $U/t$.  Results are 
obtained using exact diagonalization of matrices with linear dimensions
as large as $5.7\times 10^9$, supplemented by quantum Monte Carlo. 
We report 
the magnetic and spectral properties of C$_{20}$ as a function of $U/t$ and 
investigate electronic pair binding.
Solid forms of C$_{20}$ are studied using cluster perturbation theory and
evidence is found for a metal-insulator transition at $U\sim 4t$.
We also investigate the relevance of strong correlations to the
Jahn-Teller effect in C$_{20}$.
\end{abstract}

\pacs{71.10.Li, 71.20.Tx, 71.30.+h}

\maketitle

The smallest fullerene, C$_{20}$, contains 12 carbon pentagons 
(and no hexagons) forming a dodecahedron in a perfect 
representation of a platonic solid.
Among the many possible isomers of C$_{20}$, it is not obvious 
that this dodecahedral
fullerene cage should be the most stable and 
theoretical studies~\cite{Isomers,Galli98} have reached
different conclusions. In addition, unlike C$_{60}$,
dodecahedral C$_{20}$ is not spontaneously formed in 
condensation or cluster annealing
processes~\cite{Kroto87}, and its extreme curvature and strong reactivity 
led to doubts about its stability.
It therefore created considerable excitement when 
Prinzbach et al.~\cite{prinzbach00} 
succeeded in producing the dodecahedral fullerene isomer of C$_{20}$
in the gas-phase. Experiments have also shown evidence for
solid phases of C$_{20}$~\cite{wang01, iqbal03} although 
the crystal structure is
still debated. Density-functional studies of the solid forms 
of C$_{20}$~\cite{LDA,Miyamoto01,Spagnolatti02,iqbal03}
have suggested different crystal structures with the most promising 
candidate, C$_{20}$ cages connected by C atoms to form
a 22 atom unit cell~\cite{Spagnolatti02,iqbal03}, predicted to become superconducting upon doping with Na.\cite{Spagnolatti02} A simple-cubic-like phase of
C$_{20}$ has also been speculated to become superconducting~\cite{Miyamoto01}. In their proposal for a purely electronic mechanism
for superconductivity in C$_{60}$, Chakravarty, Gelfand and 
Kivelson~\cite{Kivelson91a} have stressed the importance of structure at 
the mesoscale~\cite{Kivelson01}.  Along with the molecular solids formed 
by C$_{60}$, solid phases of C$_{20}$ would be ideal
candidates for this picture, and a detailed understanding of 
these phases would be of great interest.
Strong correlation effects are likely to be very important in C$_{20}$, and
previous studies~\cite{Isomers,Galli98,LDA,Spagnolatti02,iqbal03,Pastor06} have treated
these correlation effects approximatively.  Here, we show that 
within a Hubbard model description, an almost exact treatment
is possible using a large-scale numerical approach.
\begin{figure}[th]
\begin{center}
\includegraphics[clip,width=\columnwidth]{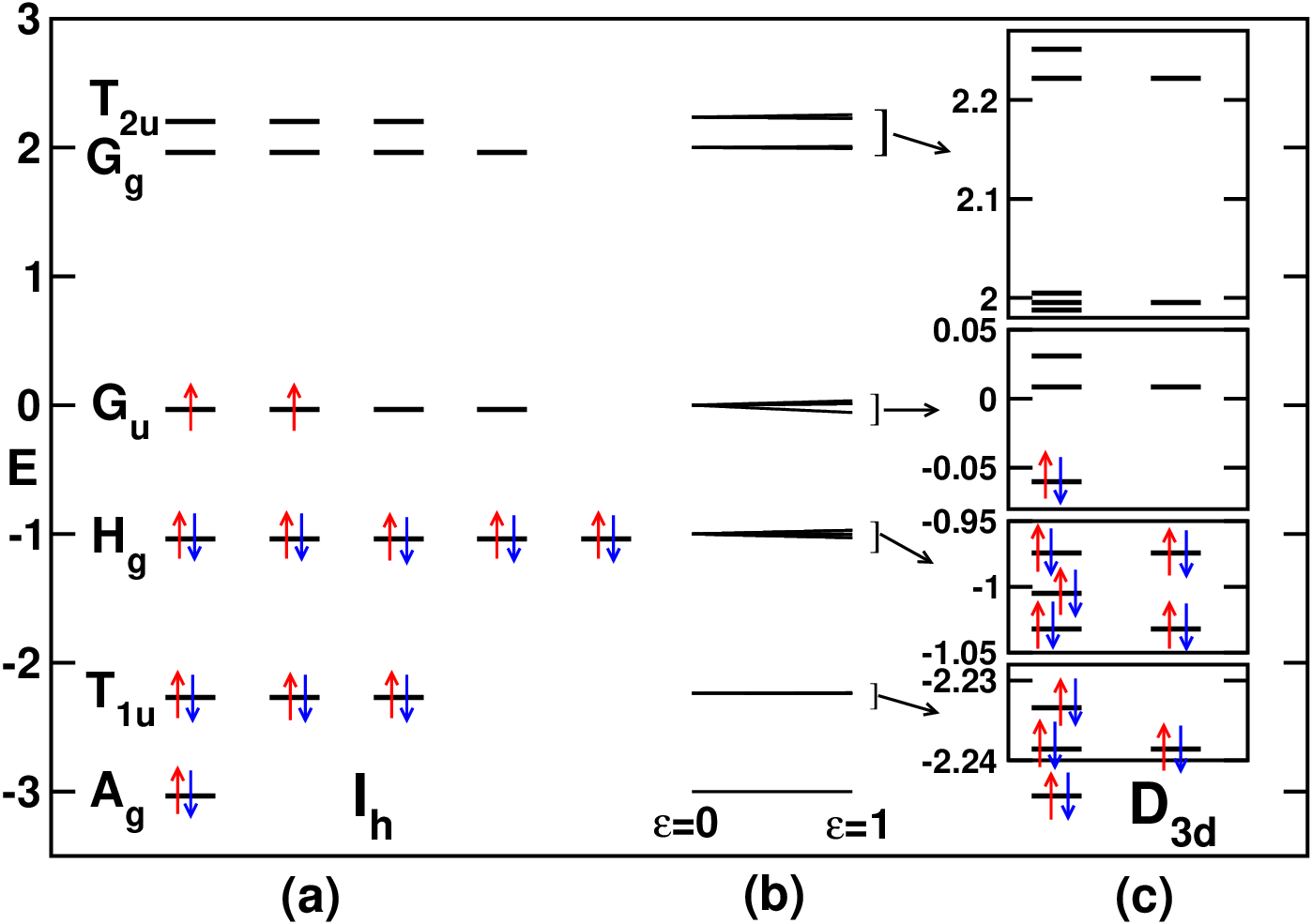}
\caption{
(Color online). (a) HMO levels for neutral 
C$_{20}$ with $I_h$ symmetry. (b) The Walsh diagram showing the
evolution of the levels with the distortion, $\varepsilon$, towards $D_{3d}$
symmetry. (c) HMO levels for the molecule with $D_{3d}$ symmetry ($\varepsilon=1$).
\label{fig:Levels}}
\end{center}
\end{figure}

Our starting point is
the one-band Hubbard Hamiltonian on a single C$_{20}$ molecule
defined as:
\begin{equation}
H=-t\sum_{\langle ij\rangle\sigma}(c^{\dagger}_{i\sigma}c_{j\sigma}+h.c.) +U\sum_i
n_{i\uparrow}n_{i\downarrow},
\label{exthubmd}
\end{equation}
where $c^{\dagger}_{i\sigma}$ ($c_{i\sigma}$) is an electron
creation (annihilation) operator,
$U$ is the on-site Coulomb
interaction and $n_{i\sigma}$ is the number of electrons on
site $i$ with spin $\sigma$. Due to the extreme curvature of
C$_{20}$ we expect $t$ to be smaller than typical values for C$_{60}$ while
$U$ should remain close to that of C$_{60}$. Consequently, we expect that
a realistic value for $U/t$ valid for C$_{20}$ is likely {\it larger} than 
the value of $U/t\sim 4$~\cite{c60spec} used for C$_{60}$, implying that strong correlation
effects play a more crucial role in the physics of C$_{20}$. 
Even though C$_{60}$ and C$_{20}$
share the same symmetry group, $I_h$, non-interacting C$_{20}$ is a metal while C$_{60}$ is an
insulator. This is evident from the H\"uckel molecular orbitals (HMO) shown in Fig.~\ref{fig:Levels}a. The highest
occupied molecular orbital (HOMO), G$_u$, is fourfold degenerate, containing 2 electrons for the neutral molecule.
The lowest unoccupied molecular orbital (LUMO), $G_g$, is considerably higher in energy. Here we shall mainly
be concerned with the neutral, 1- and 2-electron doped molecule 
which only involves the G$_{u}$ levels.

Due to the orbital degeneracy of the neutral C$_{20}$ molecule,
 the Jahn-Teller effect is likely to be important. 
The electronic G$_{u}$ levels can couple to the $A_g$, $G_g$ and $H_g$ Jahn-Teller phonon modes. Theoretical studies
have argued for the resulting lowered symmetry to be 
C$_2$~\cite{Zhang92}, $D_{5d}$~\cite{Parasuk91}, $C_{2h}$~\cite{Saito01},
$C_i$~\cite{Saito02}
and $D_{3d}$~\cite{Galli98,Yamamoto05}.  Here we follow
Yamamoto et al.~\cite{Yamamoto05} and assume a static deformation of $D_{3d}$ symmetry. The bond lengths for the optimal
$D_{3d}$ structure are~\cite{Yamamoto05} 
$a_{ab}=1.464$\AA, $a_{bc}=1.469$\AA, $a_{cc'}=1.519$\AA, and $a_{cc''}=1.435$\AA. (See Fig. 1 of Ref.~\onlinecite{Yamamoto05}.)
We parameterize the distortion by letting $t_a$, the hopping along
the bond of length $a$, depend on a parameter $\varepsilon$ in the following way: $t_a/t=1-\varepsilon(a-\bar a)/\bar a$.
Here, ${\bar a}=1.4712$\AA\ is the average over C$_{20}$ of the above bond lengths. 
Then $\varepsilon=0 $ and $1$ correspond to the $I_h$ and $D_{3d}$ structures, respectively.
At $\varepsilon=1$ the maximal deviation of $t_a/t$ from $1$ is {\it less} than 3.5\%.
In Fig.~\ref{fig:Levels}, the Walsh diagram for the evolution
of the $I_h$ levels with the distortion $\varepsilon$ is shown along with the HMO levels for the optimal $D_{3d}$ structure
for $\varepsilon=1$.

{\it Numerics --} Our exact diagonalization (ED) work is performed in a completely parallel fashion
using SHARCnet facilities. In addition to total particle number and total $S_z$ component
of the spin, we also exploit the S$_{10}$ (S$_6$) sub-group symmetry present in the $I_h$ ($D_{3d}$)
symmetry group. The basic element of this sub-group is a rotation of $2\pi/10$ ($2\pi/6$) around the face of a pentagon (around a vertex) combined with reflection.
We denote the corresponding pseudo-angular momenta by $j_{10}$ ($j_{6})$.
After symmetry reductions, the size of the Hilbert space at half-filling for the singlet states of the neutral molecule is 
$3,418,725,024$ for the dodecahedral $I_h$ configuration and 
$5,699,353,088$ for the $D_{3d}$ distorted configuration.
Using 64 cpu's, a Lanczos iteration for the $I_h$ ($D_{3d}$) configuration is completed in 
540 (980) seconds~\cite{exthub}. 
Dynamical properties are calculated using standard ED techniques~\cite{Gagliano87}.
Our quantum Monte Carlo (QMC) work follows standard methods~\cite{white89}
with ground-state energies obtained
at $T=0$ using projector QMC while spectral functions are obtained from 
finite temperature, $\beta=10/t$,  QMC~\cite{white89} combined with Maximum Entropy methods~\cite{MaxEnt}.

\begin{table}
  \centering
  \begin{tabular}{|l|l|cc|l|cc|}
    \hline\hline
       & $U=2t$ & $S$ & $j$ & $U=5t$ & $S$ & $j$ \\
    \hline
  & -20.5983834340 & 1 & 0,$\pm$2 &  -12.1112842959 & 0 & 5 \\
  & -20.5981592741 & 1 & $\pm$4 &  -12.0123014488 & 0 & 0,$\pm$2,$\pm$4 \\
\raisebox{1.5ex}[0pt]{I$_h$}  & -20.5920234655 & 0 & $\pm$2 &  -11.8770332831 & 1 & 0,$\pm$2 \\
  & -20.0527029539 & 0 & 5    & -11.8472120431 & 1 & $\pm$1,$\pm$3 \\
\hline\hline
  &  -20.6757641960 & 1 & 0 & -12.1204684092 & 0 & 3 \\
  &  -20.6462557924 & 1 & $\pm$2 & -12.0921677742 & 0 & 0 \\
\raisebox{1.5ex}[0pt]{D$_{3d}$}  &  -20.6166968560 & 0 & $\pm$2 & -11.9197052918 & 1 & 3 \\
  &    -20.0754248172 & 0 & 3 & -11.8974914914 & 1 & 0 \\
\hline\hline
\end{tabular}
\caption{Lowest energy levels (in units of $t$) of the neutral $C_{20}$ molecule
for $U = 2t, 5t$, labeled by spin and pseudo-angular momentum,
  for the $I_h$ ($j=j_{10}$) and $D_{3d}$ ($j=j_6$)
configurations.}\label{enlev}
\end{table}

\begin{figure}[t]
\begin{center}
\includegraphics[clip,width=\columnwidth]{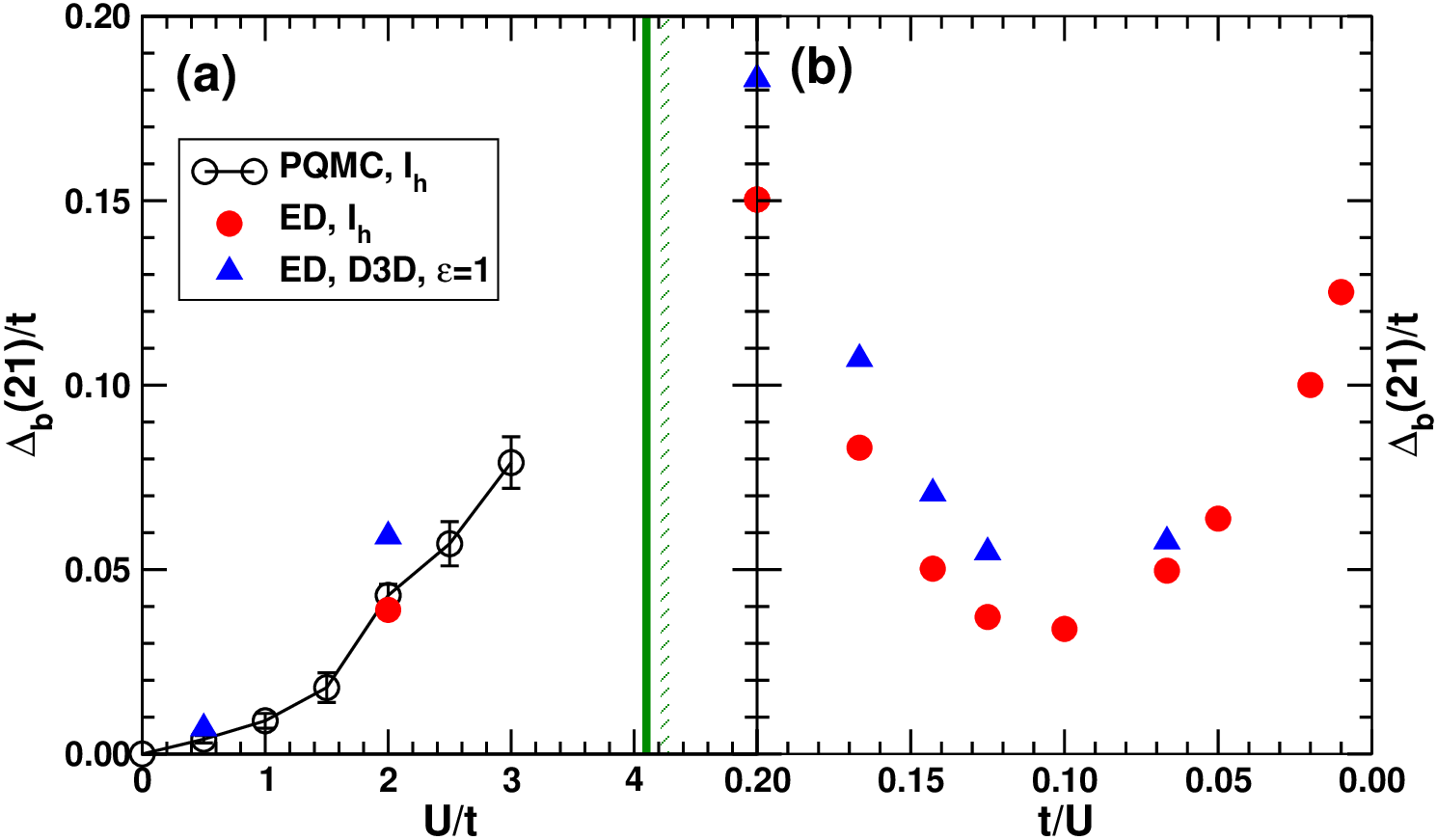}
\caption{(Color online). Electronic pair binding energy $\Delta_b(21)/t$ as a function of $U/t$.
ED ($\bullet$) and QMC ($\circ$) results for the molecule with $I_h$ symmetry. ED results ($\blacktriangle$)
for the molecule with $D_{3d}$ ($\varepsilon=1$) symmetry. 
The solid (crossed) vertical line indicates $U_c/t$
for neutral $I_{h}$ ($D_{3d}$) molecules. (a) $\Delta_b$ versus $U/t$ 
for $U/t\leq 5$. (b) $\Delta_b$ versus $t/U$ for $5\leq U/t\leq 100$.
\label{fig:Binding}}
\end{center}
\end{figure}

{\it Magnetic Properties --}
From the non-interacting HMO levels in Fig.~\ref{fig:Levels}, it would seem likely
that the ground state of the neutral molecule is magnetic at small $U/t$. Our ED and 
QMC
work confirms that this is the case for the $I_h$ configuration for $U/t\leq 3$, where the
ground state is observed to be an orbitally degenerate triplet, 
$S=1$, occurring at $j_{10}=0,\pm 2$. Table\ \ref{enlev} gives the few lowest energy levels,
labelled by spin and pseudo-angular momentum, 
for the cases of $U/t$ = 2 and 5.
For $U/t=5$, we find that the ground-state for the $I_h$ configuration 
is a non-degenerate singlet, $S=0$, occurring at $j_{10}=5$, and separated from
the lowest lying excitation, another singlet at  $j_{10}=0,\pm 2,\pm 4$, by a gap of ~0.1t.
The lowest triplet excitation is found at
$j_{10}=0,\pm 2$. This ordering of levels continues to hold for larger $U/t$,
although the energy scale decreases with increasing $U$.
The degeneracies and excitation gaps at large $U/t$ agree with ED studies 
of the dodecahedral $S=1/2$ antiferromagnetic 
Heisenberg model~\cite{Konstantinidis05}.  In fact the ground state energy
calculated for neutral $C_{20}$ in the large $U$ limit ($U > 50$) can be
related to the ED result for the Heisenberg model to the accuracy that
the latter has been calculated~\cite{ftnote}.
From Table\ \ref{enlev} it is clear that the system crosses over from a triplet to singlet ground
state between $U/t=2$ and $U/t=5$.  Assuming an approximately linear dependence on $U/t$ of 
the energy of the triplet states at $j_{10}=0,\pm 2$ and the singlet at $j_{10}=5$, we estimate
that this transition occurs at 
$U_c/t\sim 4.10$, indicated by the solid vertical line in Fig.~\ref{fig:Binding}.
The fact that the ground-state for the neutral molecule for $U>U_c$ is a non-degenerate singlet implies
that the molecule is {\it stable} against Jahn-Teller distortions.

Surprisingly, the $D_{3d}$ distorted molecule follows the same pattern with
the exception that at $U/t=0$ the unique ground-state is a singlet.
However, once $U/t$ becomes of order of the splitting of the $G_u$ levels,
$\sim 0.0686$, the ground-state becomes a triplet. At $U/t=0.5$ and $2$, we find that this ground-state
triplet occurs at $j_{6}=0$ with a number of low-lying triplet states above it.
The Jahn-Teller distortion has therefore completely removed the 
orbital degeneracy leaving only a Kramer's degeneracy.
At $U/t=2$, the lowest-lying excitation is a triplet at  $j_6=\pm 2$, and the lowest-lying singlet, with a
gap approximately twice as large, is at the same $j_6=\pm 2$.
As for the $I_h$ configuration, we observe that the ground-state is a singlet at $U/t=5$ occurring at $j_{6}=3$.  An analysis 
similar to the $I_h$ case 
yields $U_c/t\sim 4.19$, indicated by the crossed vertical line in Fig.~\ref{fig:Binding}, very close to the estimate for the $I_h$
configuration.   Again, the system remains in a singlet state for larger values of $U/t$.

{\it Pair Binding --}
The purely electronic mechanism for superconductivity~\cite{Kivelson91a,Kivelson01} is based 
on a favorable pair binding energy.
The pair binding energy is defined as the energy difference between having two extra
electrons on the same and on separate molecules:
\begin{equation}
\Delta_b(N+1) =E(N+2)-2E(N+1)+E(N).
\end{equation}
When negative, it is favorable to have the two electrons 
on the same molecule providing a purely electronic mechanism for superconductivity.
Intriguingly, for neutral C$_{12}$~\cite{Kivelson92}, as well as for
several related models~\cite{Kivelson01}, it is known that $\Delta_b$ is negative. 
For C$_{60}$,  perturbative results indicated that
the observed superconducting phase has its origins in a negative 
$\Delta_b$~\cite{Kivelson91a,Kivelson91b,Ostlund02}.
However, our earlier QMC results~\cite{lin05a} find no binding, suggesting that
either low order perturbation theory is inadequate or that the QMC results are
not sufficiently accurate to measure the binding.  Thus 
it is of considerable interest to obtain exact results for C$_{20}$ which 
can then be used to test the accuracy of QMC.

Our results for $\Delta_b(21)/t$  are shown in Fig.~\ref{fig:Binding}. 
We first consider results for the $I_h$ configuration
of C$_{20}$. The QMC results ($\circ$) and the ED results ($\bullet$) 
are in excellent agreement, and for $U/t\leq 3$ they 
show that $\Delta_b$ is positive and pair binding is {\it suppressed}. 
(For  $U>U_c$ it
is not possible to perform QMC calculations due to the sign problem.) 
The ground-state for
the neutral molecule is now a singlet and our ED results again 
clearly indicate that pair binding is {\it not} favored. As $U/t$ increases,
$\Delta_b$ reaches a minimum at $U/t\sim10$. Then, as $U/t\to\infty$,
$\Delta_b$ approaches a finite positive value, consistent 
with the exact result showing the absence of pair binding for electron doping 
in the $U=\infty$ limit~\cite{Chayes91}.
The perturbative results indicating a negative pair binding for C$_{60}$ were correlated with a violation
of Hund's rule for two electron doping~\cite{Kivelson91a}. For C$_{20}$ the same violation of 
Hund's rule occurs at large $U$. For $U/t\leq 3$,
the ground state with 22 electrons obeys Hund's rule and has $S=2$. 
However, at $U/t=5$, C$_{20}^{2-}$ has an 
intermediate spin of $S=1$ and at $U/t=6$ this state has $S=0$, fully violating Hund's rule.

\begin{figure}[t]
\includegraphics[clip,width=\columnwidth]{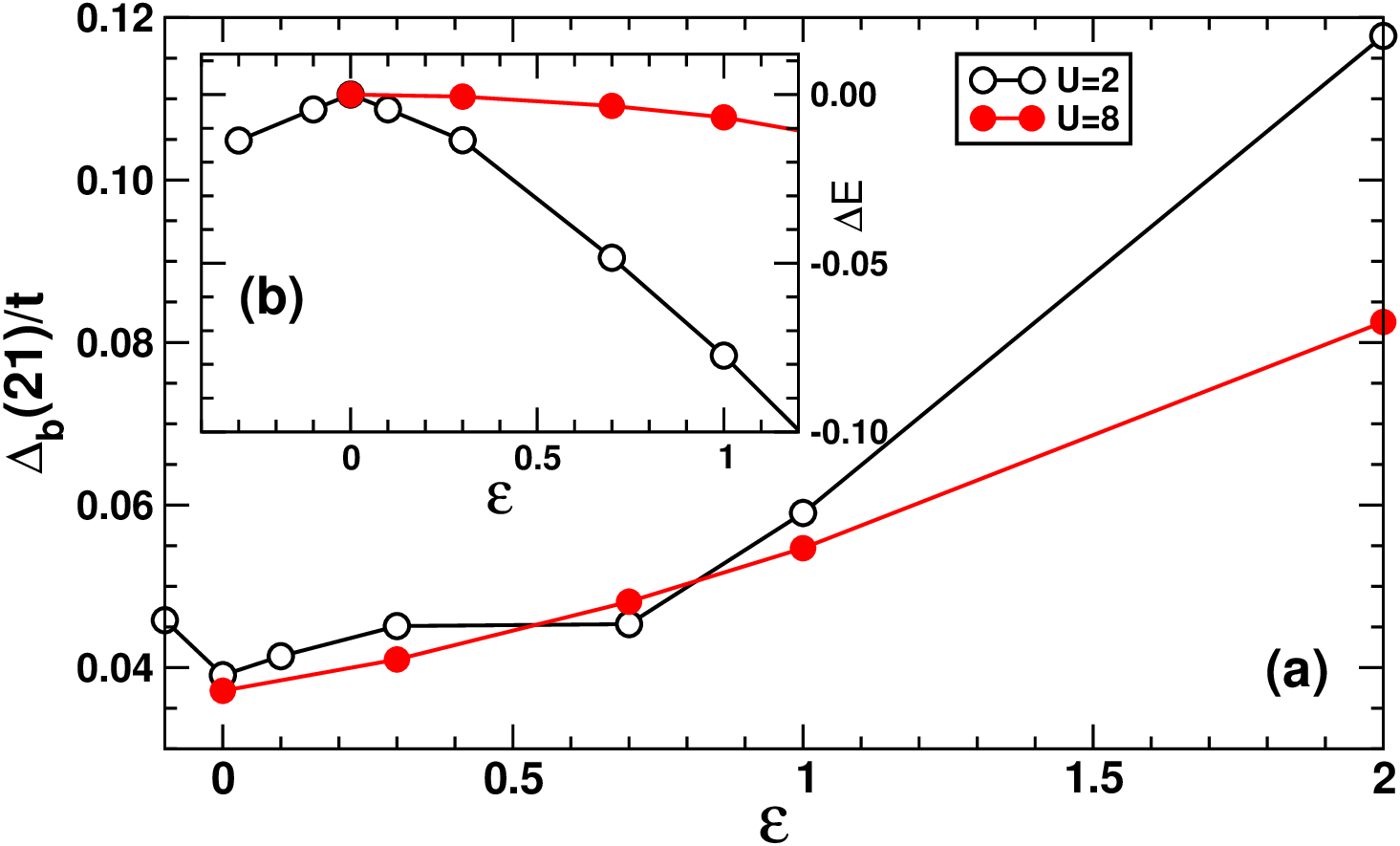}
\caption{(Color online). (a) Electronic pair binding energy $\Delta_b(21)/t$ 
versus the distortion, $\varepsilon$, towards $D_{3d}$ symmetry, for 
for $U/t=2 (\circ), 8 (\bullet)$.
The inset, (b), shows the shift in the ground state energy of the 
neutral molecule, $\Delta E$, versus $\varepsilon$.
\label{fig:Deltavseps}}
\end{figure}
We can only calculate the pair binding for the Jahn-Teller distorted molecule in an approximate manner
since the distortion will depend on the electron doping. We make the 
simplifying assumption that the
distortion is static, of $D_{3d}$ symmetry, and independent of doping. 
Since we focus on 1 and 2 electron
doping, which only involves $G_u$ orbitals, it is reasonable to assume that the symmetry of the distortion
is the same. Saito et al.~\cite{Saito02} find the same symmetry for the neutral and negatively charged molecule.
However, the optimal $\varepsilon$ will show a dependence on the doping which we ignore.
Our results for $\Delta_b$ for the $D_{3d}$ structure ($\varepsilon=1$) are shown in Fig.~\ref{fig:Binding}
($\blacktriangle$). For all values of $U/t$, we find that $\Delta_b$ for the $D_{3d}$ structure is {\it higher} than for
$I_h$. The Jahn-Teller distortion appears to work against pair binding. 
This is confirmed in
Fig.~\ref{fig:Deltavseps}a, where the pair binding energy is shown as a function of $\varepsilon$.
In Fig.~\ref{fig:Deltavseps}b, we show results for the shift in the 
ground-state energy versus $\varepsilon$ where the strong dependence on $\varepsilon$ for $U/t=2$ is indicative of the molecule
being Jahn-Teller active. By contrast,
for $U/t=8$ the dependence on $\varepsilon$ is very
shallow, signalling that the $I_h$ structure is stable.

\begin{figure}[t]
\begin{center}
\includegraphics[clip,width=\columnwidth]{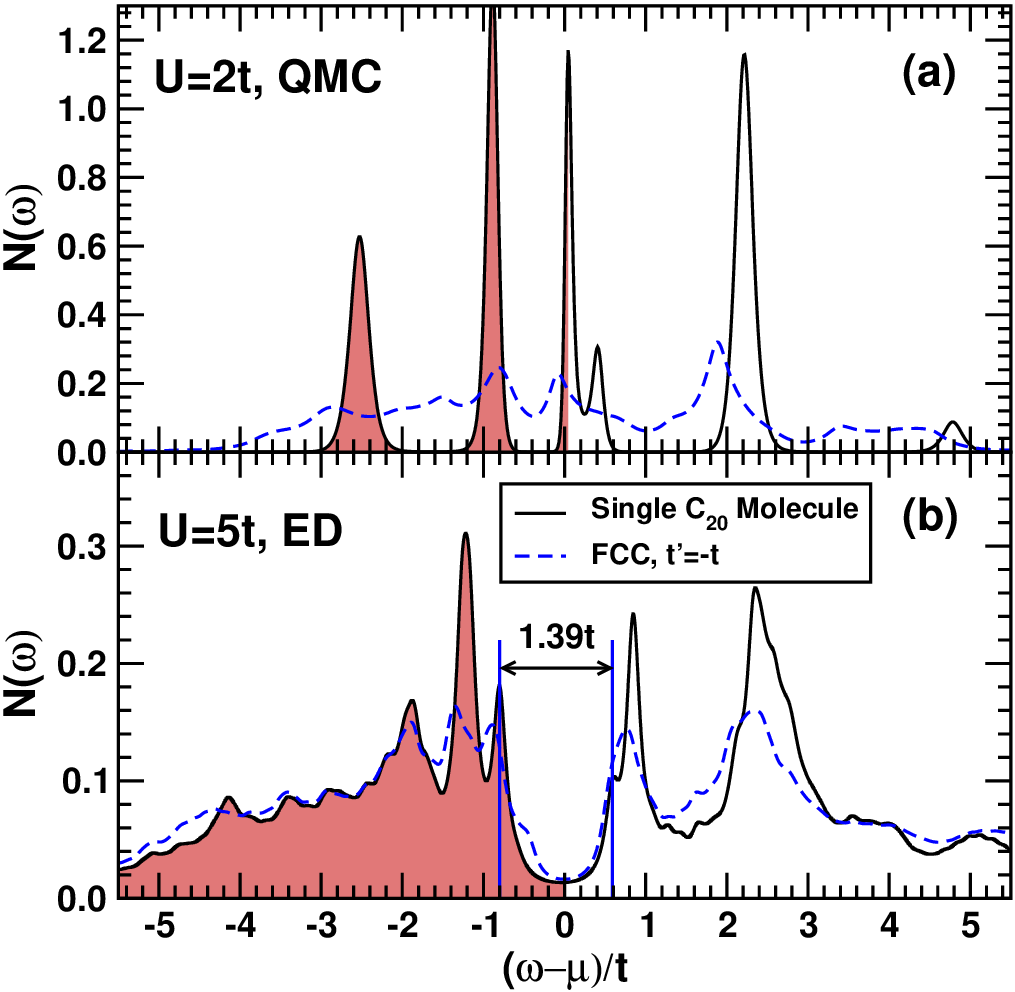}
\caption{(Color online). Density of states, $N(\omega)$, at $U=2t$ $(a)$ from QMC and at $U=5t$ $(b)$ from ED. In both panels
the DOS for the C$_{22}$ fcc lattice obtained from CPT with $t'=-t$ is shown as a dashed line.
\label{fig:dos}}
\end{center}
\end{figure}

\begin{figure}[t]
\begin{center}
\includegraphics[clip,width=\columnwidth]{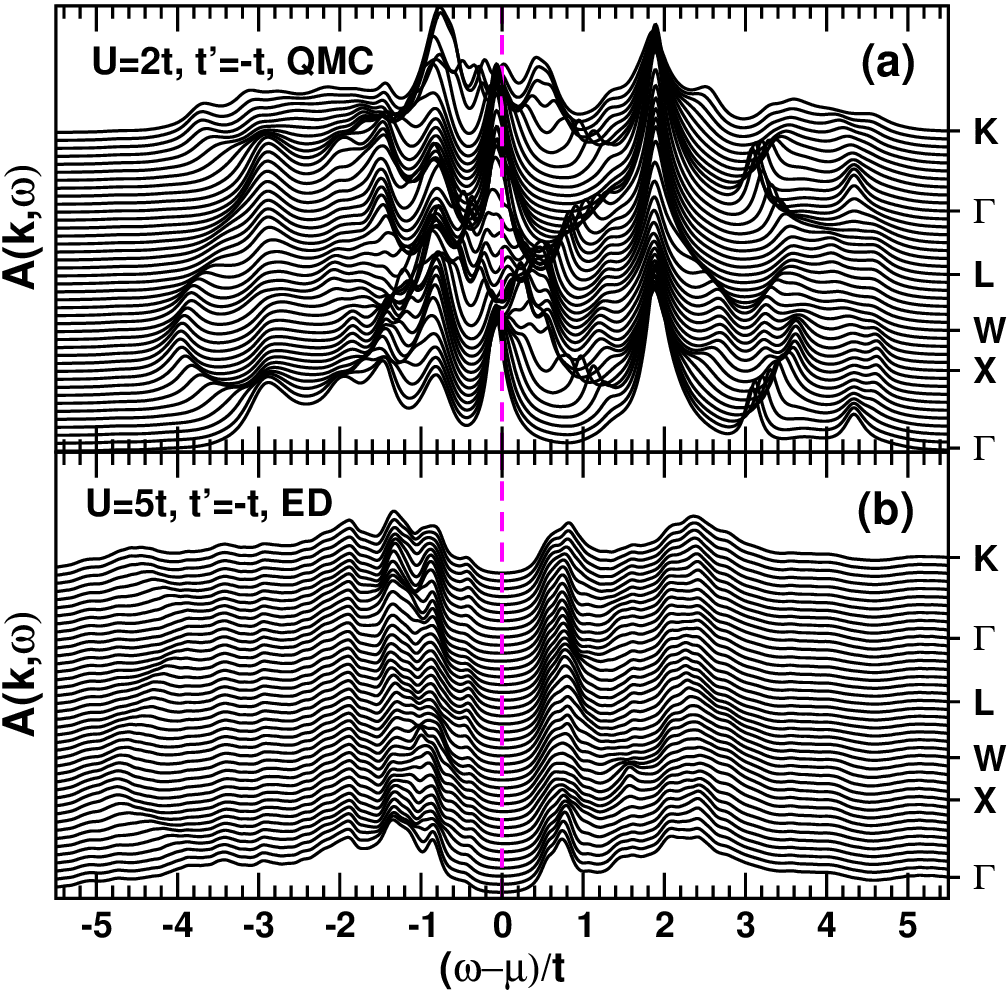}
\caption{(Color online). Spectral functions with the chemical potential 
$\mu$ determined
 from $N(\omega)$ in Fig.~\ref{fig:dos}. Results for the C$_{22}$ FCC lattice 
with $U=2t$ (a) and $U=5t$ (b) as obtained
from CPT with $t'=-t$. The k-axis labels refer to standard notation for the FCC Brillouin zone. 
\label{fig:spectral}}
\end{center}
\end{figure}

{\it Spectral Functions, Solid C$_{20}$ --}
Next we calculate the density of states, $N(\omega)$, and wave-vector dependent spectral 
functions, $A(k,\omega)=-(1/\pi){\rm Im}[G(k,\omega+E_0+ i\eta)]$, for a three-dimensional solid of C$_{20}$
molecules. 
Here, $E_0$ is the ground-state energy
and $G$ the single-particle Green's function. 
The calculation is performed by cluster perturbation theory 
(CPT)~\cite{senechal00,senechal02} using QMC and ED
data. In all cases, delta-functions were treated as Lorentzians 
with a broadening of $\eta=0.1$. The QMC version of this method was applied earlier 
to the case of C$_{60}$ monolayers~\cite {c60spec}.  We idealize the 
hypothetical fcc C$_{22}$ structure~\cite{Spagnolatti02,iqbal03}, by a 
model in which the bridging C atoms are replaced by effective hopping 
integrals, $t'=-t$, between C$_{20}$ molecules.  The resulting $N(\omega)$ and
$A(k,\omega)$, are shown in Figs.~\ref{fig:dos} and 
\ref{fig:spectral} for $U/t=2$ and 5. Also shown in Fig.~\ref{fig:dos} 
are the single-molecule densities of states.  For 
$U/t=2$, which lies in the middle of the region of triplet ground state, the 
solid of undistorted $I_h$ molecules is metallic with a complicated Fermi 
surface, while, for $U/t=5$, in the region of singlet ground state, the 
solid is insulating with a gap of about $1.4t$.

In conclusion, we have calculated the ground state properties and spectral 
functions of the Hubbard model on a $C_{20}$ molecule using 
ED and QMC.  
We have identified a ground state 
crossing at $U_c\sim 4.1t$ where the system switches from a triplet state 
which is unstable against a Jahn-Teller distortion to a gapped singlet 
state which is stable.  
Extending this result using CPT, we
identify a metal-insulator transition for the bulk solid at $U=U_c$.
If the 
symmetric, $I_h$, form of $C_{20}$ is found to be stable, a possible 
explanation would be that it is stabilized by correlations resulting from 
$U > U_c$. Additional results will be presented elsewhere~\cite{exthub}.

This project was supported by the NSERC, CIAR and CFI.
FL is supported by the US Department of Energy under award number DE-FG52-06NA26170.
AJB and CK gratefully acknowledge the hospitality and support of the 
Stanford Institute for Theoretical Physics, where part of this work was
carried out.  All the calculations were carried out at SHARCNET
supercomputing facilities.


\begin{thebibliography}{99}
\bibitem{ftnote}For large $U/t$, the neutral molecule ground state energy can be                                                                                      
    fit to $E_{20}/(t^2/U)=a+b(t/U)^2$, where $a=4E_H/J-30$ and                                                                                           
    $E_H$ is the ground state energy of the Heisenberg model,                                                                                             
    obtained by exact diagonalization in Ref.~\onlinecite{Konstantinidis05}, $E_H=-9.72219J$.  A fit, using ED results for $U/t=50$ and $100$, gives  $E_H=-9.72217J$.
\end{thebibliography}
\end{document}